\documentclass[a4paper, 11pt, notitlepage]{scrartcl}
\usepackage[margin=1in]{geometry}                % See geometry.pdf to learn the layout options. There are lots.
\usepackage[parfill]{parskip}    % Activate to begin paragraphs with an empty line rather than an indent
\usepackage[T1]{fontenc}
\usepackage{libertine}

\usepackage{amsmath}
\usepackage{amssymb}
\usepackage{graphicx}
\usepackage{setspace}
\usepackage[comma,authoryear]{natbib}
\usepackage{xcolor}
\usepackage{authblk}

\newcommand{\be}{\begin{equation}}
\newcommand{\ee}{\end{equation}}
\newcommand{\bes}{\begin{eqnarray}}
\newcommand{\ees}{\end{eqnarray}}

\title{\bf The Information Theory of Individuality}

\date{}                                           % Activate to display a given date or no date

\author[1,3]{David Krakauer}
\author[2]{Nils Bertschinger}
\author[2]{Eckehard Olbrich}
\author[2,3]{Nihat Ay}
\author[1,3]{Jessica C. Flack}

\affil[1]{Wisconsin Institute for Discovery, University of Wisconsin, Madison, USA}
\affil[2]{Max Planck Institute for Mathematics in the Sciences, Leipzig, Germany}
\affil[3]{Santa Fe Institute, Santa Fe, USA}

\begin{document}

\maketitle
%\section{} 
%\subsection{}

\begin{abstract}

\noindent We consider biological individuality in terms of information theoretic and graphical principles. Our purpose is to extract through an algorithmic decomposition system-environment boundaries supporting individuality. We infer  or detect evolved individuals rather than assume that they exist. Given a set of consistent measurements over time, we discover a coarse-grained or quantized description on a system, inducing partitions (which can be nested). Legitimate individual partitions will propagate information from the past into the future, whereas spurious aggregations will not. Individuals are therefore defined in terms of ongoing, bounded information processing units rather than lists of static features or conventional replication-based definitions which tend to fail in the case of cultural change. One virtue of this approach is that it could expand the scope of what we consider adaptive or biological phenomena, particularly in the microscopic and macroscopic regimes of molecular and social phenomena. 


\end{abstract}

\section*{The architecture of individuality}

From the perspective of physics and chemistry, biological life is surprising. There is no physical or chemical theory from which we can derive biology, and yet if we break down any biological system into its elementary constituents there is no chemistry or physics that remains unaccounted for \citep{GellMann:1995uh}. The fact that physics and chemistry are universal -- ongoing in stars, solar systems and galaxies -- whereas to the best of our knowledge biology is exclusively terrestrial, supports the view that life is physically emergent and perhaps unique. This stands in contrast to the universality of chemical phenomena which can be derived from quantum mechanical considerations in fundamental physics, even when this proves to be computationally cumbersome or intractable \citep{Defranceschi:2000tn,Schaefer}.  The asymmetry in what can be gleaned from working down towards ever more elementary constituents versus working up through levels of aggregation is captured in the terms reductionism and emergence. It is the emergent property of biology --which implies mechanistic explanatory gaps between biochemistry and cell biology, cell biology and individuals, and individuals and societies -- that has proven so challenging. The emergence of individuals in particular is one of those features of biology that has attracted considerable attention, and forms the basis of many areas of specialized biological knowledge, from taxonomy and cladistics through to physiology, behavior, and ecology \citep{Clarke,Wilson13}.  

It is almost inconceivable for us to imagine a biological science without a concept of individuality. After all, how could we speak about metabolism, behavior or the genome without first establishing a unit or container of observation and measurement? Even Schr\"{o}dinger in his  prescient book, 
{\em What is Life?} \citep{Schrodinger:2012vd}, that sought to explore the persistence of biological phenotypes of organisms -- or even features of ecosystems -- through the lens of  elementary and universal physical underpinnings, made strong prior assumptions about the reality of individual organisms:

``What degree of permanence do we encounter in hereditary properties and what must we therefore attribute to the material structures which carry them? The answer to this can really be given without any special investigation. The mere fact that we speak of hereditary properties indicates that we recognize the permanence to be of the almost absolute. For we must not forget that what is passed on by the parent to the child is not just this or that peculiarity, a hooked nose, short fingers, a tendency to rheumatism, haemophilia, dichromasy, etc. Such features we may conveniently select for studying the laws of heredity. But actually it is the whole (four- dimensional) pattern of the phenotype,  all the visible and manifest nature of the individual, which is reproduced without appreciable change for generations, permanent within centuries -- though not within tens of thousands of years -- and borne at each transmission by the material in a structure of the nuclei of the two cells which unite to form the fertilized egg cell. That is a marvel.''

It is clear that Schr\"odinger did not set out to derive the individual from fundamental physics, but to reconcile existing and rather traditional conceptions of individuality (essentially the individual as synonymous with the observable organism) with the new physics of quantum mechanics. 

In this respect Schr\"odinger was adopting a typically reductionist perspective, explaining features of biological science through first principles of physics \citep{Weinberg:1995wz}. In Schr\"odinger's case, the physical feature of greatest importance to biology was the long-lived covalent bond. But for many reasons this line of approach has failed to deliver the deep and unifying insights based on physics \citep{Anderson:1972wi}, from which powerful biological ideas -- such as adaptation or individuality -- might be derived \citep{Dupre09,Keller:09}.

The question that we seek to address in this chapter is more modest. We are not asking how biology emerges from physics and chemistry \citep{Rasmussen:2004wo}, but how we might go about identifying individuals through an automatic procedure by tracking  the emergence of new spatial and temporal aggregations and their coordinated and adaptive properties. We  propose that all evolutionary processes must leave tell-tale signatures of coordinated patterns of activity among putatively unrelated processes -- that these coordinated behaviors are frequently  recognized as a form of individuality (for a synoptic treatment of these see \citep{Clarke}).  Significantly elevated coordination is a feature of multicellular organisms as well as microbes and loosely-bound ecological assemblages such as microbial mats. In an ideal case, visitors to an exoplanet would have a procedure for identifying individuals based on a quantitative survey with minimal prior knowledge of the type of life forms that they expect to encounter. Individuality is therefore a signature of life.

We suggest that a productive research program in individuality should allow \textit{a-priori} for the possibility of multiple degrees of individuality at all levels of biological organization. This is in contrast to assigning individuality to a single level or object --  such as replicating cells or organisms -- and then applying \textit{post hoc} biological definitions such as sequestered germ cells, vertical transmission of genetic material, a common pool of metabolic free energy, or coordinated immune responses. It is not that these features are not effective criteria when we have significant prior knowledge of a system. It is that they are often derivative of the space-time properties that support individuals and fail to apply at super-organismal levels. 

We further propose that a desirable property for a model of individuality is that it will be able to account for the hierarchical organization of life into nested trophic and functional levels -- multiple parallel levels of individuality. We take this position to be related to the recent suggestion of \citep{Rieppel} where he argues for individuals based on hierarchical complexes of homeostatic properties. 

Our technical objective is to allow researchers to extract  natural levels from dynamical systems, determining -- through a principled, algorithmic decomposition -- provisional system-environment boundaries, and then  measuring and quantifying each subsystem's degree of individuality. The extent to which we find evidence of individuality becomes a measure of the action of an evolutionary or adaptive process. 

The approach is initiated with microscopic measurements of some extended dynamical system as suggested in the abstract. This could be a vector of chemical concentrations over time, the abundance of various cell types, or the probabilities of observing dependable behaviors. These provide a list of consistent measurements on a number of traits of possible relevance to discerning individuality within parts of a system over the course of time. The measurements are then grouped in a principled way, coarse-grained or quantized into  a smaller number of partitions (which can be nested). Some of these partitions will continue to behave in a coordinated fashion over characteristic intervals of time. If they do so we have grounds to believe in the value of a given partition. If not then we will have incorrectly grouped observations or there is no coordinated behavior to detect.  Differences in the rates of change of partitions will provide evidence of functional separations including organism-environment boundaries -- organisms will typically change quickly relative to their environments.

These  ``natural'' time and space scales should reveal clues about the the mechanisms driving their consolidation into individuals \citep{Flack:2012da,Flack:uma}. The formal basis for this approach to aggregation comes from information theory, and throughout this paper we assume that individuals are best thought of in terms of dynamical proccesses and not as stationary objects. In this respect our approach might reasonably be framed through the lens of ``process philosophy'' \citep{Rescher} which makes the elucidation of the dynamical and coupled properties of natural phenomena the primary explanatory challenge. From the perspective of  ``process philosophy'' the tendency of starting with objects and then listing their properties -- ``substance metaphysics'' --  places the cart before the horse.

\subsection*{The biological individual}
In biology the canonical individual is often some form of replicator \citep{Wilson13} -- either a single biological organism developed from a fertilized egg, with individuality residing at the phenotypic level (see \citep{Dawkins:1983tz}), or asexual microbes or clonal organisms for whom individuality is defined based on shared genetic ancestry \citep{Hughes:1989ve}.  This pervasive replicator assumption has served as the starting point for theorizing about what an individual is in a broad class of studies. As a result, three widely accepted properties of biological individuals include (1) they can increase in relative frequency exploiting a source of metabolic free energy, (2) they respond adaptively to their environments, and (3) they are characterized by tightly coordinated relationships (chemical, physiological, computational) among their parts. Note that none of these properties require that the individual is a single organism. Cells and even societies can possess all defining features, as a number of authors have recognized, particularly in connection to a variety of mutualisms \citep{Gilbert}.

One problem with this approach is that the environment or context of the candidate individual is often neglected, and more often than not, this is treated as nothing other than all parts of the larger ecological context that are not members of the coordinated set comprising the individual. We argue that without a rigorous definition of the environment it is very difficult and likely impossible to speak consistently of individuals. This is analogous to figure-ground separation in gestalt psychology or computer vision. The background of an image carries as much if not more information than the object, and the challenge is to separate the two rather than assume that they are already distinct and independent.

In summary, almost all definitions of individuality  start from an assumption of set members (individual) and a set complement (environment). These are articulated in different ways in the various fields, including: (1) as an immunological concept pertaining to the idea of self and non-self \citep{Pradeu}, (2) as a temporal aggregate encoding a common past separable or independent from the past of other aggregates (ontogenetic or phylogenetic) \citep{Rieppel} (3) as a spatially bounded collection of metabolic reactions insulated by a membrane from reactions in the environment \citep{Rasmussen:2004wo} , and (4) as a unit of selection and evolutionary change \citep{LWB1987,Hughes:1989ve,Szathmary:1997wu,Callcott:2011wv}. Our goal is to discover dynamical set members and their complement.

\subsection*{Challenges to conceptualizing individuality}
A useful exercise in thinking about the pros and cons of one current biological view of individuality (see also, \citep{Santelices:1999tn}) is to consider objects or processes that do not always get classified as individuals, as these can reveal hidden assumptions informing our intuitions and judgement. 

Work on the social insects and on a number of plant, fungal and prokaryotic species has made clear a few of the pitfalls of this ``intuitive'' approach to the definition of individual \citep{Gow:2008tz,Esser:2001tz}. What these taxa demonstrate is that when we speak of individuals we are often talking about individuality at multiple organizational levels -- physically distinct ants, but also aggregates of individual ants forming colonies, and they are not always physically contiguous. For example, the majority of worker ants do not replicate, the colony as whole does not replicate, but replication is a feature of the system, and, importantly, the combination of replication by some colony members, with the persistence (and hard work) of other colony members, allows the colony as a whole to persist and adapt in response to changes in the environment. 

Viruses occupy a sort of twilight zone in biology. Declared by some non-living, and treated by most as a rather pathetic minimal limit of life, viruses constitute obligate translational parasites, incapable of completing their life cycles without first appropriating the protein synthesis machinery of a host cell. The viral capsid contains a largely inert genome responsible for encoding only a small fraction of the proteins required for synthesizing a new virus genome and the capsid required for egressing from the infected cell. The virus exists only within the larger dynamical, regulatory network of the cell. Hence the virus  -- understood as the active parasitic agent -- is comprised largely of host encoded factors. And yet it can replicate, adapt and has a persistent identity that distinguishes it from its environment -- despite the fact it relies on its environment for replicating. We might say that the virus is one small part of a chimerical individual \citep{Krakauer:2006ua}.

\subsection*{A way forward}
\label{forward}
One possibility is that ant colonies and viruses are only nominally individual - a property of our human perception. But if they are real in a deeper physical sense then how might we determine this? We suggest that one way forward lies in allowing \emph{a priori} for the possibility of degrees of individuality at different levels of biological organization rather than restricting ourselves to the unique individuality of the organism. In other words, we do not assume that individuality is binary but allow it to be continuous, with the surprising implication that some processes possess greater individuality than others. We also allow individuality to be nested such that individuals can contain individuals. These two properties are well accepted in biology.  Organelles (as individuals) are nested within cells (also individuals), and some cells act more independently than others -- blood cells versus neurons. Nevertheless these desirable properties have not formed the basis for a theory that  allows us to identify individuals when we have only limited prior knowledge of a system.

\section*{Quantifying individuality}

We seek a means of extracting sets of causally correlated observations reflecting coordinated biological activity through an automated or algorithmic procedure. Presented with a set of time varying measurements, we follow a procedure for partitioning these into sets of variables that represent some plausible definition of individuality based on non-trivial temporal coherence. We provide a procedure that is sufficiently tolerant of variation in underlying mechanism such that a variety of ``biological individuality'' concepts are respected. 

For static structures, there already exist reasonable definitions of modularity. Many of these definitions are associated with procedures for partitioning microscopic data into tightly bound groups, such as communities. For example, in networks quantitative modularity measures seek partitions of nodes and edges into sets that are statistically over-represented in data when compared to an appropriate null model \citep{NewmanMEJ:2006}. Unlike notions of individuality, modules are usually derived from static or equilibrium systems. For an individual, we desire a dynamical system in order to include not only a separability concept (for say self non-self distinctions) but also an historical dimension and a mechanism of self-preservation or long-term stability in the sense of Schr\"odinger . 

Developmental definitions of modularity, such as those applied to limb formation, or the appearance of body segments, also provide a window into individuality \citep{Davidson04} but they have not been presented in the form of algorithmic methods for automated extraction and identification as they have in network science \citep{NewmanMEJ}. In this paper we are restricting ourselves to these ``automated'' or ``algorithmic'' procedures based on a belief in the value of mechanizing a concept and implementing it on a suitable programmable machine. In this respect we are following the mathematician and computer scientist, Alan Turing's approach to defining computation in terms of Turing machines operating on Turing tapes \citep{Turing}. This is an operational or instrumental approach in which a concept such as computation or individuality, comes to be understood in terms of a procedure with well defined input and output properties. One of Turing's epistemological innovations was to replace the somewhat contentious library of competing definitions of computation with one very general formalism of conceptually synthetic value and also of utilitarian value (this paper was written on a jazzed-up Turing machine).

\subsection*{The origin of information} 
A brief foray into the origin of the concept of information is required to understand the basis for the approach we develop for quantifying the degree of individuality a system exhibits.

The roots of the information theoretic interpretation of entropy lie in the physical theory of thermodynamics and the formal definition of work introduced by Clausius in the 1860s (see \citep{Muller:2007ul} for an introduction to this history).  Work (displacement of a physical system) is produced by transferring thermal energy from one body to another (heat). Entropy captures, or measures, the loss in temperature over the range of motion of the working body.  In other words, entropy measures the energy lost from the total available energy available for performing work. The insights of Clausius were formalized and placed in a mathematical framework by Gibbs in 1876.

In 1877 Boltzmann provided an alternative interpretation of entropy in his kinetic theory of gases as a measure of the potential disorder in a system. This definition no longer emphasizes energy dissipated through work but the number of unobservable configurations (microstates) of a system, e.g. particle velocities consistent with an observable measurement (macrostate), e.g. temperature.  These definitions of entropy are closely related as Boltzmann entropy increases following the loss of  energy available for work attendant upon the collision of particles in motion during heat flow. There are many different microscopic configurations of individual particles compatible with the same macroscopic measurement, only a few of which are useful.

In 1948, encouraged by John von Neumann, Claude Shannon used the thermodynamical term entropy to capture the information capacity of a communication channel. A string of a given length (macrostate) is compatible with a large number of different sequences of symbols (microstates). A target word  will be disordered during transmission in proportion to the noise in a channel. If there were no noise, each and every microstate could be resolved and the entropy would define an upper limit on the number of signals that could be transmitted.  The study of the maximum number of states that can be transmitted from one point to another across a channel, in the face of noise and when efficiently encoded is called information theory.

Shannon did not describe entropy in terms of heat flow and work, but information shared through a channel transmitted from a signaler to a receiver \citep{Shannon}. The power of information theory derives in part from the incredible generality of Shannon's scheme. The signaler can be a  phone in Madison and the receiver a phone in Madrid, or the signaler can be a parent and the receiver its offspring. For phones the channel is a fiber-optic cable and the signal pulses of light. For organisms the channel is the germ line and the signal the sequence of DNA or RNA polynucleotides in the genome. Increasing entropy for a phone-call corresponds to the loss or disruption of light-pulses, whereas increasing entropy during inheritance corresponds to mutation or developmental noise. The same scheme can be applied to development, in which case the signaler is an organism in the past and the receiver the same organism in the future. One way in which we might check that we are dealing with the same organism is that the information transmitted forward in time is close to maximal.

In its simplest form, Shannon made use of the following formal measures when defining information. The entropy $H$ of a random variable $S$ measures the uncertainty or information of the states that it can adopt:
	
\begin{eqnarray*}
  H(S) = -\sum_i P(s_i)\log_2 P(s_i)
\end{eqnarray*}
where $s_i$ are the possible values of the state and $P(s_i)$ the probabilities of these states. For a coin there would be two possible values for $S$, heads and tails, and the values of these states for a fair coin would be the probability $0.5$, yielding a metric entropy value of $1$. Deviation from a fair coin corresponds to a reduction in information, as in the limit of bias where only one side of the coin is favored, the outcome is known in advance and any toss of the coin is perfectly predictable. This produces an entropy value of $0$. Hence information is minimized when predictability is maximized. 

To capture the communication value of information Shannon introduced a signaler and receiver structure, which is now typically described using two random variables $S$ and $R$. The maximum information transmitted between signaller and receiver is given by the Mutual Information ($I$). The $I$ can be written in several different forms. One intuitive expression is:

\begin{eqnarray*}
  I(S;R) = H(S) + H(R) - H(S,R)
\end{eqnarray*}
Where $H(S)$ and $H(R)$ are the entropies of the signals, and $H(S;R)$ the joint entropy of the two variables,

\begin{eqnarray*}
  H(S;R) = -\sum_i \sum_j P(s_i, r_j)\log_2 P(s_i, r_j)
\end{eqnarray*}
The joint entropy is at a maximum when there is no relationship between the $S$ and $R$ variables. The $I$ is therefore high when the information in $S$ and $R$ are high and they are strongly coupled in their values  ($H(S;R)$ is low). The $I$ measures the information shared between $S$ and $R$ over a communication channel because the only source of structure in $R$ is assumed to come from $S$. 

Another popular way of writing $I$ is,
\begin{eqnarray*}
  I(S;R) = H(R) - H(R|S)
\end{eqnarray*}
where $H(S|R)$ is the conditional entropy of $R$ or the amount of information in $R$ that is not in $S$. Hence if all the information in $R$ comes from $S$ then $H(R|S)$ will be zero, and $I(S;R) = H(R)$. 

These measures capture a maximum of shared information and provide statistics for an informational theory of the individual. 

\subsection*{The informational individual} 

For purposes of this paper, we begin with the assumption that biological individuality can be usefully understood in terms of an informational individual. This is not to be confused with Dawkin's replicator, as replication might be inessential to individuality. What is essential is the idea that information can be propogated forwarded in time, meaning that uncertainty can be reduced over the course of time. 

What follows is an attempt to capture coordinated, persistent and stable structures or regularities in an information theoretic language in order to quantify an hypothetical system's degree of individuality. It is our belief that when biologists speak of individuals they are often invoking informational individuals without always making this assumption explicit. In this way biological individuals can be seen to be a natural extension of the ideas of Boltzmann and Von Neumann, originating in ideas related to statistical mechanics and thermodynamics. This helps to build a bridge to statistical physics without reducing or claiming to explain all the levels at which interesting or improbable biological structure and function are observed. Our approach is closely related to ideas associated with the concept of autopoiesis developed by Maturana \citep{Maturana:1975ge} who emphasizes the ``unity'' of a network of processes engaged in self-production in terms of autonomy, and the application of autonomy to problems of cognition \citep{Maturana:1980vx}. 

\begin{enumerate}

\item {\it The system environment decomposition}. Consider a dynamical set of quantifiable measurements that we coarse grain into components of a system and components of an environment. We seek a way of establishing whether this partition is justifiable, and whether the individuality concept is relevant.  We wish to allow for a hierarchy of such partitions in order to capture biological examples such as organelles within cells, and cells within bodies within populations, where in each case the target entity and the environment assume a different identity. Here we will simplify the problem a bit by assuming the organism is the environment of the cell and the population is the environment of the individual organism. The exercise will be to pick arbitrary partitions, and then apply a suitable criterion for invoking the individuality concept, retaining only those partitions that meet strict inclusion criteria.

\item {\it Informational individuals}. In the pursuit of generality, we consider a discrete, stochastic dynamical system, where the state of the system in the future is determined by some subset of states in the present. If we arbitrarily divide these states into system and environment, we should like to be able to determine how the current system state $S_n$ and the current state of the environment $E_n$ together are sufficient to determine the next system state $S_{n+1}$. Formally the predictability of the next state of the system is quantified via the mutual information:
\[  I(S_n, E_n; S_{n+1}) = H(S_{n+1}) - H(S_{n+1} | S_n, E_n)  \]
This expression seeks to capture how much information at time $n+1$ $S_{n+1}$ comes from the system itself at a previous time step (or generation) $S_n$ -- versus from the environment at a previous time $E_n$. 
This mutual information can now be decomposed in two ways
\begin{eqnarray*}
  I(S_n, E_n; S_{n+1}) & = & I(S_{n+1}; S_n) + I(S_{n+1}; E_n | S_n) \\
			& = & I(S_{n+1}; E_n) + I(S_{n+1}; S_n | E_n)
\end{eqnarray*}
Each decomposition can be interpreted as different ways of distributing the observed past regularities between the system and environment\footnote{There are attempts \citep{Williams2010,Harder2013,Bertschinger2013} to come up with a finer decomposition of the mutual information that would allow to determine which information is shared by the system and the environment, which information is unique to the system and environment, respectively, and a complementary or synergistic part. Such a decomposition would resolve the ambiguity in the attribution to either the system or the environment as expressed by the two interpretations as genomic or environmental determination.}. Each of these allow us to define different forms of individuality. 

\begin{enumerate}
\item {\it Genomic determination}.   Consider $I(S_{n+1}; S_n) + I(S_{n+1}; E_n | S_n)$: \\
Here we consider the influences of the system state onto itself (at the next generation or time step). For a preferred interval of time, all observed dependencies between successive system states are attributed to the system only. The environment state appears only in so far as the environment is influenced by prior states of the system. In this case, the future is only a function of intrinsic mechanisms of heredity in the system and not maintained through interaction with the environment.
The quantity $I(S_{n+1}; S_n)$ has been called autonomy $A^*$. It should be high when the system is in control of its environment and thus can determine its next state effectively. \\

The remaining influence of the environment, as measured by $I(S_{n+1}; E_n | S_n)$, can be interpreted as new information for the system flowing from the environment into the system. It is new to the system in the sense that it cannot already be predicted from the state of the system. When this information flow vanishes, a system can be said to be {\em informationally closed}. So this quantity measures the degree to which the system is not closed $nC$. Note that closure does not require causal independence, it only states that all influences from the environment are predictable by the system. 

\item {\it Environmental determination}. An alternative to genomic determination is structure imposed by the environment.  Consider $I(S_{n+1}; E_n) + I(S_{n+1}; S_n | E_n)$: \\
Here the observed influences are attributed to the environment (as far as possible according to $I(S_{n+1}; E_n)$).  Only the remaining influence $I(S_{n+1}; S_n | E_n)$ is thought to be due to the system. This can be interpreted as an alternative source of {\em  autonomy} $A$ for the system. It should be valid under the assumption that all dependencies between the states of the system and the environment are attributed to the environment. 
\end{enumerate}

Note that even if this is not suggested by the intuition behind the two  measures A and $A^*$, it is possible that $A^* < A$.
Such a situation has been called {\em complementary} (or synergistic) since it means that neither the system nor the environment alone predict much about the next system state. Instead their joint action is required to determine $S_{n+1}$. A standard example of this situation would be the XOR-function in boolean logic. \\[2ex]
Thus altogether the following relationship is obtained between the measures of autonomy, dependency and closure
\[  A^* + nC = A + I(S_{n+1}; E_n)  \]
Interpreting $I(S_{n+1}; E_n) - nC$ as a measure of non-trivial closure, since it is large if the system has information about its environment, but still achieves informational closure (low $nC$),
one obtains
\[  A^* - A = NTIC   \quad  \mbox{($NTIC$ = non-trivial informational closure)}  \]

\subsection*{Reflections on Closure and sufficiency}

We can think about an individual as a system partition that is a sufficient predictor of its own future. This means in particular that $S_{n-1}$ does not add any information about $S_{n+1}$ besides the one already contained in $S_n$. Formally this reads as $I(S_{n+}; S_{n-1} | S_n) = 0$.  \\
Note that by the Markovian structure of the system environment interaction
\begin{eqnarray*}
I(S_{n+1}; S_n, E_n) & = & I(S_{n+1}; S_{n-1}, S_n, E_n) \\
			& = & I(S_{n+1}; S_n) + [ I(S_{n+1}; S_{n-1} | S_n) + I(S_{n+1}; E_n | S_{n-1}, S_n) ] \\
			& = & I(S_{n+1}; S_n) + I(S_{n+1}; E_n | S_n)
\end{eqnarray*}
Thus $I(S_{n+1}; E_n | S_n) = I(S_{n+1}; S_{n-1} | S_n) + I(S_{n+1}; E_n | S_{n-1}, S_n)$ which means that informational closure implies sufficiency, i.e.
\[  nC=I(S_{n+1}; E_n | S_n) = 0  \implies  I(S_{n+1}; S_{n-1} | S_n) = 0 \]
Informational closure is therefore a stronger notion than sufficiency\footnote{For a more general setting this was shown in \citep{Pfante2014}.} which allows the system
to be influenced by the environment as long as this influence cannot be predicted from within the system. \\[2ex]

\subsubsection*{Considering longer histories}

The above calculations can be generalized to longer histories:
\begin{eqnarray*}
I(S_{n+1}; S_n, E_n) & = & I(S_{n+1}; S_{n-l}^n, E_{n-k}^n) \\
			& = & I(S_{n+1}; E_{n-k}^n) + I(S_{n+1}; S_{n-l}^n | E_{n-k}^n) \\
			& = & I(S_{n+1}; S_{n-l}^n) + I(S_{n+1}; E_{n-k}^n | S_{n-l}^n) \\
			& = & I(S_{n+1}; S_{n-l}^n) + [ I(S_{n+1}; E_n | S_{n-l}^n) 
				+ \underbrace{I(S_{n+1}; E_{n-k}^{n-1} | S_{n-l}^n, E_n)}_{= 0} ]
\end{eqnarray*}
Thus overall the same relationships between autonomy, closure and sufficiency are obtained
\begin{eqnarray*}
  I(S_{n+1}; E_{n-k}^n) + \underbrace{I(S_{n+1}; S_n | E_{n-k}^n)}_{A_k} & = &
  		\underbrace{I(S_{n+1}; S_n)}_{A^*} + \underbrace{I(S_{n+1}; S_{n-l}^{n-1} | S_n)}_{\mbox{(non-)sufficiency}} 
+ I(S_{n+1}; E_n | S_{n-l}^n) \\
% 	& = &	A^* + \mbox{ (non-)sufficiency } + I(S_{n+1}; E_n | S_{n-l}^n) \\
	& = & 	A^* + nC
\end{eqnarray*}

\subsection*{Algorithmic identification of individual boundaries} 

We have provided formal algorithms that capture an important evolutionary concept of individuality. We now use these algorithms to identify individuals in their environments. The basic idea is to systematically increase the number of variables that we assign to the target system and determine whether this procedure leads to an increase in the quantity representing autonomy. If the expansion of the boundary of the system does not lead to an increase in autonomy, then we have incorporated an environmental variable needlessly. In this way individuals represent mechanism for aggregating dynamical processes in such a way as to maximize their knowledge of the future.  We can implement this procedure using both the autonomy and non-closure quantities. If autonomy increases as we expand our system and non-closure decreases, then we have grounds for the belief that we are capturing more of the individual by including more processes formerly treated as environmental.

Let us denote the original system with $S$ and the part of the initial environment which becomes the system by $\Delta S$. The remaining environment should be denoted by $E'$. 
\begin{itemize}
\item{\bf Informational closure} \\
Then we have the two information flows
\begin{eqnarray*}
nC&=&I(S_{n+1};E_n|S_n) \\
  &=&I(S_{n+1};E'_n \Delta S|S_n) 
\end{eqnarray*}
and
\[
nC'=I(S_{n+1} \Delta S_{n+1};E'_n| \Delta S_n,S_n) 
\]
Using some algebra we get
\[
nC'=nC-I(S_{n+1};\Delta S_n|S_n)+I(\Delta S_{n+1};E'_n|\Delta S_n, S_{n+1},S_n)
\]
The first term subtracts the information flow which is now internalized and the second term adds the flow which resided previously in the environment. Clearly the system becomes more closed if the internalized flow is larger than one included additionally. 
\item{\bf Autonomy} \\
Let us start with the simpler measure $A^*$, the mutual information between subsequent states:
\[
A^*=I(S_{n+1};S_n)
\]
and 
\begin{eqnarray*}
A'^*&=&I(S_{n+1} \Delta S_{n+1};S_n \Delta S_n) \\
&=&A^*+I(\Delta S_{n+1};S_n|S_{n+1})+I(S_{n+1} \Delta S_{n+1};\Delta S_n|S_n) \;.
\end{eqnarray*}
Similarly one obtains for the other dependency quantity:
\begin{eqnarray*}
A&=&I(S_{n+1};S_n|E_n) \\
&=&I(S_{n+1};S_n|E'_n \Delta S_n)
\end{eqnarray*}

\begin{eqnarray*}
A'&=&I(S_{n+1} \Delta S_{n+1};S_n \Delta S_{n}|E'_n) \\
&=&A+I(S_{n+1} \Delta S_{n+1};\Delta S_n|E'_n)+I(\Delta S_{n+1};S_n|E'_n \Delta S_n S_{n+1}) \\ 
&=&A+I(S_{n+1};\Delta S_n|E'_n)+I(\Delta S_{n+1};S_n \Delta S_n|E'_n S_{n+1}) 
\end{eqnarray*}
Both autonomy measures can only grow or stay constant with increasing system size, but they can never decrease. 

\subsubsection*{Sufficiency expansion and boundary detection}

We can explore sufficiency a little more formally in terms of Markov conditions.
We consider a system as self-sufficient if its dynamics are Markovian, i.e. 
\[
I(S_{n+1};S_{n-m}|S_{n-1},\ldots,S_{n-m+1})=0 \;.
\]
In the following we consider only the case $m=1$:

We get for the non-sufficiency $nS'=I(S'_{n+1};S'_{n-1}|S'_n)$ of the enlarged system 
\be
I(S'_{n+1};S'_{n-1}|S'_n)=I(S_{n+1} \Delta S_{n+1};S_{n-1} \Delta S_{n-1}|S_n \Delta S_n) 
\ee
Now we apply the chain rule:
\begin{eqnarray*}
I(S_{n+1} \Delta S_{n+1};S_{n-1} \Delta S_{n-1}|S_n \Delta S_n)&=&I(S_{n+1} \Delta S_{n+1};S_{n-1}|S_n \Delta S_n) \nonumber \\&& +I(S_{n+1} \Delta S_{n+1};\Delta S_{n-1}|S_{n-1}, S_n \Delta S_n) \\
I(S_{n+1} \Delta S_{n+1};S_{n-1}|S_n \Delta S_n)&=&I(S_{n+1};S_{n-1}|S_n \Delta S_n)+I(\Delta S_{n+1};S_{n-1}|S_n \Delta S_n S_{n+1}) \\
I(S_{n+1} \Delta S_{n+1};\Delta S_{n-1}|S_{n-1}, S_n \Delta S_n)&=&I(S_{n+1};\Delta S_{n-1}|S_{n-1}, S_n \Delta S_n) \\ && +I(\Delta S_{n+1};\Delta S_{n-1}|S_{n-1} S_n \Delta S_n S_{n+1}) \\
\mbox{and} && \\
I(S_{n+1};S_{n-1}|S_n \Delta S_n)&=&I(S_{n+1};S_{n-1}|S_n)+I(S_{n+1};\Delta S_n|S_{n-1} S_n) \\ && -I(S_{n+1};\Delta S_{n} |S_n)
\end{eqnarray*}
which leads to 
\begin{eqnarray*}
nS' &=& nS-I(S_{n+1};\Delta S_{n} |S_n)+I(S_{n+1};\Delta S_n|S_{n-1} S_n)+I(S_{n+1};\Delta S_{n-1}|S_{n-1}, S_n \Delta S_n) \\ && +I(\Delta S_{n+1};\Delta S_{n-1}|S_{n-1} S_n \Delta S_n S_{n+1})+I(\Delta S_{n+1};S_{n-1}|S_n \Delta S_n S_{n+1}) \\
nS'&=&nS-[I(S_{n+1};\Delta S_{n} |S_n)-I(S_{n+1};\Delta S_n|S_{n-1} S_n)] \\&&+I(S'_{n+1};\Delta S_{n-1}|S_{n+1} S'_n)+I(\Delta S_{n+1};S_{n-1}|S'_n S_{n+1})
\end{eqnarray*}
We see that the only term which can lead to a decrease of the non-sufficiency is $I(S_{n+1};\Delta S_{n} |S_n)-I(S_{n+1};\Delta S_n|S_{n-1} S_n)$. It can be interpreted as the  internalized information from $S_{n-1}$ to $S_{n+1}$. \\
The other two terms  are always positive and related to new information flows through the environment -- which are made possible by enlarging the system and which are not accounted for by $nS$.  
\end{itemize}
\end{enumerate}

\subsection*{Interpretation of information theory of individuality}

Using an information theoretic framework applied to a time series we have derived a number of principled quantities that capture degrees of individuality. The exposition is neccessarily formal as we have sought to provide a procedure for ``discovering'' individuals in a variety of different biological contexts. The key empirical requirement is the careful measurement of a number of hypothesized individual attributes or properties  over the course of time. For example, the abundance of organisms in a population; the genetic or phenotypic states of cells or tissues over time, the firing rates of neurons over time. In each case we require a consistent time series of measurements in an appropriate coordinate frame (concentration, spatial position, firing rate, chemical concentration) that provide the input to our algorithms. It is our contention that many existing biological concepts (e.g.tightly coordinated replicators, developmental individuals), will be identified through this procedure. Many novel ``individuals'' will also be identified, including those at the societal level that are currently deprecated as derivative or epiphenomenal.

 We derived a quantity for autonomy, $A$ which measures to what degree the past of a system can predict its future. Whether we are speaking of the persistence of a cell, an organism, or a personality-type, we have in mind finding that unit of description able to account for some property of that system in the future. We even use the terms ``individualistic" to refer to relative independence from the environment. We only need to consider what a low autonomy individual entity would look like -- uncoordinated in space and time, with parts that do not work together -- to see how critical this value becomes.  If systems with high autonomy are what is meant by individuals, its complement, non-closure $nC$, when high suggests that we have found the wrong unit of analysis. We are not dealing with an individual but a collection of poorly integrated components. This is because the only way to explain a system with a high value of $nC$ is to invoke everything outside of the system, as this value is precisely the environmental contribution to a system. 

There are occasions however where some complex combination of the system and the environment is required, and this is the basis of our measure $A^* - A$ or non-trivial information closure. This is expected to be high when the individual and the environment enter into complex synergistic relationships. It is perhaps the notion of individuality best suited to communities of organisms. Each organism will obtain a high value of $A$ whereas the community or colony will obtain a high value of $A^* - A$. In this way these two measures allow us to distinguish between the individual in the environment versus individuals nested within super-individuals.

These individuality-related quantities suggest a natural procedure for identifying the boundaries of the individual. By incrementally titrating the system/environment boundary (inclusion and exclusion criteria) we expand the individual to be the largest set of processes which upon further addition is no better at predicting its future than the previous largest set. For example, if we included not only the motor system, but metabolism and behavior in our definition of an individual organism, our ability to project its behavior through time would be significantly improved. If we  included the limbs of a second individual or the metabolism of its preferred diet species in the individual, our prediction of the organism is unlikely to increase.  This is not to say that the environment is not important, only that the environment is not the same as the individual. If the distinction cannot be made, this would be a very intriguing instance of a non-individualistic dynamical system, as it would imply that there is no favored grouping of processes, only a random distribution of events through time. 

The choice of information theory as a favored theoretical framework allows for great generality but forfeits mechanism. In this sense the information theoretic theory of individuals is rather like the Darwinian selection theory that does not rest on any one mechanism of selection, but seeks to explain how selection can yield skewed distributions of genotypes and phenotypes through time. Both are formal schemes for interpreting a diverse body of data, and seek to provide a coherent framework for analysis and explanation. An advantage of this approach is that it can provide through logic a common foundation for discussing a range of nominally distinct observations, and tracing a path from biology to probability theory and statistical mechanics, helping to build connectivity among biotic and abiotic phenomena. 

Hence the more traditional discussions of membranes, self versus non-self, symbiogenesis and the like have not been treated explicitly. All of these approaches provide the physical basis for the functional requirements that we have sought to present. Although the theory has nothing to say about mechanism it is important to realize that it is explicitly motivated by an understanding of the role that mechanism plays in the consolidation of functional space and time scales and emergence of new function in biological systems, as discussed in Sec. \ref{forward} and in Refs \citep{Gintis:2012uo,Flack:2012da,Flack:uma}

\section*{Caveats and challenges}

The purpose of this paper has been to place the discussion of biological individuality on a solid logical and probabilistic foundation. In order to do so we have taken a fair amount for granted, including the ability to make accurate measurements at arbitrary scales of granularity and over scales of time that are biologically meaningful. We have also neglected to discuss  those mechanisms that make heredity possible in the first place, in other words, robustness mechanisms that enable the error-free, or low error, transmission of information across generations. We have avoided discussing the specifics of the functional or selective benefits of hierarchical levels, concentrating on their identification. It is fair to assume that long lived aggregates could develop the capacity to replicate and become a significant target for selection and hence a bona fide level at which selection operates, which for some is what is implied by biological individuality \citep{Okasha}, in which case our approach could provide a means of identifying both pre-individuals (low autonomy) as well as fully-fledged individuals (high autonomy). We discuss each of these topics in more detail below. 

\begin{enumerate}

\item {\it The partitioning requirement}. In the previous discussion we have `defined' the quantities, autonomy, closure and sufficiency, in terms of system and environment, but we have not provided any mechanisms that might generate a time series with appropriate values, or discussed how we might go about identifying the best system and environmental variables in the first place. Moreover the choice of time scale will be instrumentally critical, as over very short or very long time scales we are unlikely to observe the regularities from which we seek to derive the individuality quantities: autonomy, sufficiency and closure. It is our belief that few of these attributes (system and environment variables, time scales etc) can be known in advance, and that it is precisely through the algorithmic determination of the individual that each will obtain relative support. 

\item {\it The robustness requirement}. Further to identifying nested or hierarchical partitions, we also require some specification of the machine itself - the generator of the time series. This will be equated with parts of the individual and needs to possess some level of robustness or an error-correcting property. This is because individuality in adaptive systems often seems to be associated with adaptive mechanisms of homeostasis --  mechanisms that monitor internal states and ensure that deviations are minimized. It is this self-preserving quality of the individual that allows us to make some useful discrimination between physical phenomena and biological ones, without exaggerating the dynamical differences. 

\item {\it The levels of selection}. In many previous treatments of individuality, the idea that the individual has a special evolutionary status has been posited. This is presented in terms of levels of selection, where coarse-grained aggregates achieve a coordinated persistence property that now allows them to be treated as segregating, selective units. The most popular formalism for thinking about this process is presented in  terms of the Price equation, which describes how the mean value of a trait changes as a function of the covariance in that trait and fitness, and the previous value of the trait. Of interest to us here is that the Price equation assumes some partition of trait values into groups and attempts to do this in such a way as to best capture the evolution of the mean value of the trait in the population. Assuming some true underlying structure and dynamics, the accuracy of the equation will depend on the choice of partition \citep{Krakauer:2010uk}, and our information theoretic approach could provide such a principled platform for modeling.

\end{enumerate}

\section*{Functional implications}

Consider the observation that from genomes, to cells, tissues, individuals, societies, and eco-systems, evolution generates structures with nested spatial and temporal levels (\emph{e.g.}, \citep{Feldman:1982tp,LWB1987,Smith:1997wr,Valentine:1996ud,Michod:2000uh,Gould:2002ts,Feldman:1982tp,Jablonka:2005vq}). Often with each new structural level there arises new functionality: a new feature with positive payoff consequences. This new functionality can be present in the form of a new behavioral output such as a feeding response to a previously inaccessible resource.  

Elsewhere we have suggested that a primary driver for the emergence of new functions is the reduction of environmental uncertainty through the construction of dynamical processes with a range of characteristic time constants, described using nested slow variables. Slow variables are coarse-grained encodings of fast, microscopic dynamics. Proteins, for example, have a long persistence relative to RNA transcripts, and can be thought of as the summed output of translation. Cells have a long persistence relative to protein turnover rates and are a function of the summed output of arrays of spatially structured proteins. Both proteins and cells encode some average quantity derived from the noisier activity of their constituents. Slow variables are useful because they are informative about the future state of the system by capturing the regular patterns from the past. Hence an appropriate average protein density will be more informative about the rate of gene expression and RNA translation than any individual measurement of transcription or translation. 

As a consequence of integrating over abundant microscopic processes, slow variables provide better predictors of the local future configuration of a system than the states of the fluctuating microscopic components. As long as there are regularities in the microscopic dynamics, these can be detected by the system components, and the statistics describing these regularities are available to be computed by the components when possible. Slow variables can reduce environmental uncertainty and by increasing predictability promote accelerated rates of microscopic adaptation. The reduced uncertainty facilitates adaptation in two ways: It allows components to fine-tune their behavior and it frees components to search at relatively low cost a larger space of strategies for obtaining resources from the environment. A related phenomenon has been studied  in relation to neutral networks arising from RNA folding. Many different sequences can fold into the same secondary structure. This implies that over time, structure changes more slowly than sequence, thereby freeing sequences to explore many configurations under normalizing selection \citep{Fontana:1998uv}.

We suggest that \emph{individuality} becomes a functionally significant concept when: (1) some system's components rely to a greater extent on the coarse-grained features of dynamics for adaptive decision-making than on the detailed and often noisy full microscopic dynamics. And (2) when other processes or systems interacting with the target system encode only this coarse-grained behavior, or slowly changing coordinated behavior of the target system, and ``fail'' to see the target system's microscopic behavior. Ultimately an individual is a statistically quantized unit of prediction and thereby the elementary  constituent of all adaptation.  

\section*{Aknowledgements. }

 We thank Lynn Nyhart and Scott Lidgard for their motivating questions, guidance and careful thoughts on this topic. We thank Cormac McCarthy for his close reading and recommendations. JCF and DCK were supported by the U.S. Army Research Office MURI award under contract number W911NF-13-1-0340 and the John Templeton Foundations for their support through The Principles of Complexity award to the Santa Fe Institute.

\bibliographystyle{chicago} 
\bibliography{Individuality}

\end{document}